      [1999/12/01 v1.4c Il Nuovo Cimento]
\documentclass{cimento}
\usepackage{epsfig}

\newcommand\be{\begin{eqnarray}}
\newcommand\ee{\end{eqnarray}}
\newcommand\nn{\nonumber}

\title{``Stopping'' of light and quantum memories for photons}
\author{M.~Fleischhauer and C. Mewes}
\instlist{\inst{} Fachbereich Physik, Universit\"at
Kaiserslautern, E.-Schr\"odinger Str.46, D-67663 Kaiserslautern, Germany}
\PACSes{
\PACSit{42.50.-p}{},
\PACSit{42.50.Gy}{}, 
\PACSit{42.65.Tg}{}, 
\PACSit{03.67.-a}{}}

\begin{document}

\maketitle

\begin{abstract}
A coherent technique for the control of photon 
propagation in optically thick media 
and its application for quantum memories is discussed. 
Raman adiabatic passage with an externally controlled Stokes field can be used
to transfer the quantum state of a light pulse (``flying'' qubit) 
to a collective spin-excitation (stationary qubit) and thereby slow down its
propagation velocity to zero. 
The process is reversible and has a potential fidelity of unity without the 
necessity for strongly coupling resonators. A simple quasi-particle
picture (dark-state polariton) of the transfer is presented. The analytic 
theory is supplemented with exact numerical solutions.
Finally the influence of decoherence mechanisms on 
collective storage states, which are $N$-particle 
entangled states, is analyzed.
\end{abstract}


\section{Introduction}


Among the many challenges for the implementation of quantum
information processing (QIP) 
is the transport of unknown quantum states
between separated locations as well
as the realization of quantum memories with short access times 
\cite{DiVincenzo00}.
Quantum optical systems are very attractive for this, since
photons provide ideal carriers of quantum information and
nuclear-spin or hyperfine states of atoms are ideal storage 
systems. Isolation from environmental interactions on one hand
and controllable coupling to the photon field on the other can  
most easily be realized through Raman transitions with a classical
laser providing the Stokes field.  Combining Raman coupling with 
adiabatic following
 provides control over the 
coherent absorption or emission of photons. 

The application of Stimulated Raman adiabatic passage (STIRAP) \cite{STIRAP} 
to single-atom cavity
systems has lead to a number of important proposals for qubit transfer 
between atoms and photons as well as for quantum-logic  gates
\cite{cavity-QED}. 
However due to the small absorption cross section of an 
isolated atom, it is necessary to employ strongly coupling resonators. 
The realization of the strong-coupling regime for optical frequencies 
remains a technically very challenging 
task despite the enormous experimental progress in this field 
\cite{cavity-QED-prob}.
Furthermore a single-atom system is by construction highly 
susceptible to the loss of atoms and requires a high degree of
control over atomic positions. For these reasons 
the prospects of cavity-QED systems in qubit transfer and storage
may be limited.

On the other hand a photon is absorbed with certainty, if 
a sufficiently large number of atoms is present.  
Normally such absorption is accompanied  
by {\it dissipation} and only a partial mapping of quantum properties
of light to atomic ensembles can be achieved \cite{Polzik-old}. 
If dissipation is involved, it is in general not possible 
to store the quantum state of photons 
on the level of {\it individual} quanta (single qubits) in a reversible 
manner.
Rather a quasi-stationary source is required, whose output can be 
considered as a train of wave-packets in identical quantum states.
Similar limitations apply to techniques of classical optical-data 
storage in the time domain based on spin- and photon echo or Raman
photon echo \cite{photon_echo}.

Recently we have proposed a method that
combines the enhancement of the absorption cross section
in many-atom systems  with dissipation-free adiabatic passage
\cite{Fl00-pol}. 
It is based on Raman adiabatic 
transfer of the quantum state of photons to {\it collective atomic 
excitations} using electromagnetically induced transparency (EIT) \cite{EIT}. 

In EIT a strong coherent Stokes
fields renders the otherwise optically thick medium transparent
for a resonant pump. 
Associated with the
transparency is a large linear dispersion, which has been demonstrated 
to lead to a substantial reduction of the group velocity 
of light \cite{group}. When a pump pulse propagates
in such a medium, its front end gets coherently ``absorbed''
 and the corresponding
excitation is transferred to the Stokes field as well as 
into a spin excitation.
At the back end the process is exactly reversed and all excitation
returned to the field, i.e. there is no net transfer
from photons to atoms or vice versa. From the point of view of the atoms
slow light in an EIT systems corresponds to a complete adiabatic return.
Nevertheless part of the photonic excitation is temporarily stored 
in the atoms. During this time interval
photons and atoms form quasi-particles, called
dark-state polaritons \cite{Fl00-pol} that propagate with a velocity
determined by the ratio of the electromagnetic to the matter component. 
Unfortunately EIT systems have only limited capabilities
as a temporary memory, since the achievable 
ratio of storage time to pulse length is rather limited \cite{Hau99b}.

The limitations of EIT can be overcome, however, when the group 
velocity is changed {\it in time}
and is adiabatically reduced to zero by dynamically decreasing
 the strength of the
classical Stokes field. The reduction of the Stokes field in time
brakes the symmetry of the process and
 net flow
of excitation from photons to atoms or vice versa can be achieved.
By adiabatically reducing the group velocity 
the light pulse can eventually be brought to a full stop.  In this
process its excitation
as well as its quantum state
are completely transferred to the atomic spins. 
The process can be reversed and the light pulse regenerated. 
Recent experiments \cite{stop_exp} have already 
demonstrated some basic principles of this technique - the dynamic 
group velocity reduction to a full stop and adiabatic 
following in dark-state polaritons. 


\section{single-atom cavity-QED}


To introduce the basic idea of a controlled
and reversible quantum state transfer via Raman adiabatic passage, let us
first consider the case of an individual 3-level atom coupled to
a single quantized mode as pump and a classical field as
Stokes field. This is illustrated in Fig.~1. Both fields are assumed
to be resonant with the corresponding transitions and decay from the excited
state out of the system is taken into account. 
The (complex) interaction ``Hamiltonian''
in rotating-wave approximation reads in a rotating frame
\begin{equation}
H = \hbar g a\, {\sigma}_{ab} + 
\hbar\Omega(t)
{\sigma}_{ac} + {\rm h.c.} -i\hbar\gamma {\sigma}_{aa} 
\label{ham0}
\end{equation}
where $a, a^\dagger$ are the annihilation and creation operators
of the quantized pump mode and $\Omega(t)$ is the (in general time-dependent)
Rabi-frequency of the classical Stokes field. $\sigma_{ij}\equiv 
|i\rangle\langle j|$ is the atomic spin-flip operator 
from state $j$ to state $i$
and $g=\wp\sqrt{\omega/\hbar\epsilon_0 V}$
characterizes the vacuum Rabi-frequency of the quantized mode, 
$\wp$ being the dipole moment
and $V$ the mode volume.
The imaginary part of the Hamiltonian effectively describes spontaneous
emission from the excited state with rate $\gamma$. 
The interaction couples only triplets of bare eigenstates of the combined
atom-field system, viz $|b,n+1\rangle \leftrightarrow |a,n\rangle
\leftrightarrow |c,n\rangle$, where $n=0,1,\dots$ denotes the number of photons
in the  mode. In addition there is the total ground state
$|b,0\rangle$ which is completely decoupled. In the basis of these states
the Hamiltonian separates in $3\times3$ block-matrices of the form
\begin{equation}
{\sf H}_n=\hbar\left[\matrix{-i\gamma  & g\sqrt{n} & \Omega(t)\cr
                           g\sqrt{n} & 0 & 0\cr
                           \Omega(t) & 0 & 0\cr}\right].
\end{equation}
%
%

\

\begin{figure}[ht]
\centerline{\epsfig{file=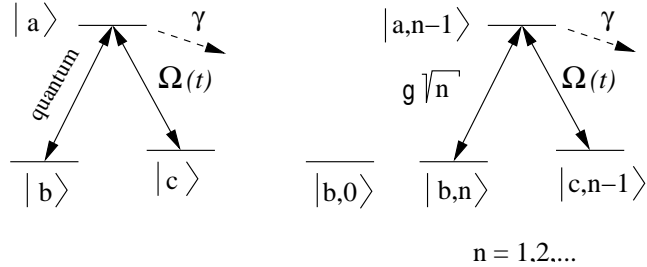,width=8.5 cm}}
\vspace*{2ex}
\caption{a) 3-level atoms coupled to single quantized mode and classical
control field of (real) Rabi-frequency $\Omega(t)$. b) coupling of relevant
bare eigenstates; $g$-vacuum Rabi-frequency}
\label{3-level}
\end{figure}

\


\noindent 
Each of these matrices has one instantaneous (adiabatic) eigenstate
$|\phi_0^{(n)}\rangle$ with eigenvalue $\epsilon_0^{(n)}=0$ and two 
eigenstates $|\phi_{\pm}^{(n)}\rangle$ with eigenvalues $\epsilon_\pm^{(n)}=
\pm \sqrt{\Omega^2(t)+g^2 n}$. The 
zero eigenstates read
\begin{eqnarray}
|\phi_0^{(n)}\rangle = \cos\theta_n(t)\, |b,n+1\rangle -\sin\theta_n(t)\,
|c,n\rangle, \qquad n=0,1,2,\dots,
\end{eqnarray}
where the mixing angles $\theta_n(t)$ are defined as
$\tan\theta_n(t) \equiv {g\sqrt{n}}/{\Omega(t)}$.
The important feature of the zero-eigenstates is that they do not
contain the excited state and are thus immune to spontaneous emission. 
For this reason these states are called dark states \cite{dark}.
Furthermore by changing the strength of the classical Stokes field, i.e.
$\Omega(t)$, the mixing angles $\theta_n$ can be rotated from $\theta_n=0$,
where $|\phi_0^{(n)}\rangle = |b,n+1\rangle$ to $\theta_n=\pi/2$, where
$|\phi_0^{(n)}\rangle = -|c,n\rangle$. If the
atom-field system is initially prepared in the dark state and if
$\Omega(t)$ 
changes sufficiently slowly, the state vector will follow 
the rotation.
Thus  by stimulated
Raman adiabatic passage \cite{STIRAP} 
a controlled and reversible transfer of excitation from the 
quantized radiation mode to the atom is possible:
\begin{equation}
|b,n+1\rangle\, \longleftrightarrow\, -|c,n\rangle.
\end{equation}
This mechanism is the basis of several proposals for engineering of
quantum states of the radiation field in resonators, for the 
transfer of quantum states between atoms through 
a resonator mode and for the transfer of quantum states between
different cavities \cite{cavity-QED}.

Adiabatic following requires that the rate of change of the mixing
angles should be sufficiently slow. In particular the characteristic
time of transfer $T$ should obey the condition
\begin{equation}
\frac{g^2 n}{\gamma} \, T \, \gg 1.
\end{equation}
The transfer time $T$ is usually limited by the finite decoherence
time of the field. In a resonator set-up the lifetime is determined
for example by mirror losses and coupling to the outside.
If $\kappa$ denotes the characteristic rate of photon 
loss, the decoherence time of a Fock-state
$|n\rangle$ is $1/(n\kappa)$. Thus adiabaticity requires
\begin{equation}
g^2 \gg \kappa\, \gamma\label{strong-coupling}.
\end{equation}
This condition is referred to as strong-coupling regime.
Condition (\ref{strong-coupling}) is technically
very difficult to satisfy. The physical origin of this strong
condition is the small absorption cross
section $\sigma$ of atoms in the optical frequency domain. 
In fact as $g^2 \sim \sigma/A\sim\lambda^2/A$, where $A$ is the cross section
of the light field at the position of the atoms, tight focusing is required.
As a consequence cavity-QED techniques
with individual atoms in the strong-coupling regime
are very sensitive to an exact
positioning of the atom. Resonator systems have the further disadvantage
that communication between them requires a careful timing of the control
fields to avoid losses due to impedance mismatch at the cavity interfaces.


\section{Temporary storage of photons in many-atom systems:
electromagnetically induced transparency and slow-light}


An obvious way to overcome the limitations of single-atom systems 
caused by the small optical cross section is to use optically thick
ensembles of atoms. This also allows to abandon the use 
of resonators and thus to avoid
impedance matching problems. In order to identify appropriate
 transfer schemes in ensembles of 3-level atoms, let us first
discuss pulse propagation in such
media. 

One of the most important phenomena associated with pulse propagation
in these systems
is called electromagnetically induced transparency 
\cite{EIT}. If, as indicated in Fig.~2, a sufficiently strong coherent 
field of constant Rabi-frequency $\Omega$ couples the excited state $|a\rangle$ to a meta-stable state $|c\rangle$,
the absorption of 
the probe field  is exactly canceled at 
two-photon resonance. In Fig.~2 the imaginary part of the
susceptibility 
\begin{eqnarray}
\chi=\eta\, \frac{\gamma\delta}
{|\Omega|^2 -\delta^2 -i\gamma \delta}
\approx  \eta\left[\frac{\gamma\delta}{|\Omega|^2} +i
\left(\frac{\gamma\delta}{|\Omega|^2}\right)^2  +{\cal O}(\delta^3)\right]
\end{eqnarray}
is shown as a function of the 
detuning from resonance.
Here $\eta=\frac{3}{8\pi^2} \rho\lambda^3$ is a density parameter 
that gives the number of atoms per cubic wavelength, $\delta$ is the
probe detuning and $\gamma$ the excited-state decay rate.


\

\begin{figure}[ht]
\centerline{\epsfig{file=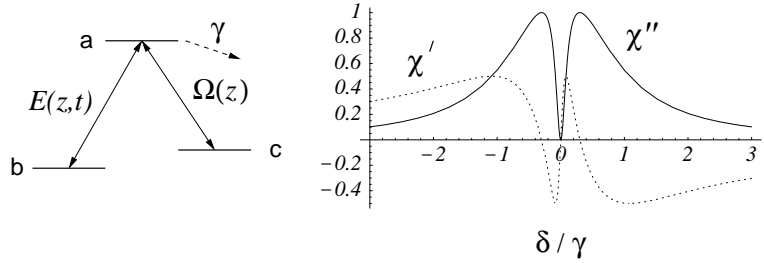,width=10.0cm}}
\vspace*{2ex}
\caption{{\it left:} 3-level $\Lambda$-type medium resonantly coupled 
to a coherent Stokes
field with Rabi-frequency $\Omega$ and a (quantum) probe field $E(z,t)$.
{\it right:} Typical susceptibility spectrum for probe field $E$ 
as function of normalized detuning 
for resonant and constant drive field. 
Real part $\chi^\prime$ describes refractive index
contribution and imaginary part $\chi^{\prime\prime}$ absorption.
}
\label{1-d}
\end{figure}

 At the same time the real part $\chi^\prime$, 
which is responsible for the refractive index, shows a large normal linear 
dispersion. 
Associated with the linear dispersion is a reduction of the group velocity
\begin{eqnarray}
v_{\rm gr}=\frac{c}{1+n_g(z)},\qquad{\rm with}\qquad
n_g(z)=\omega\frac{{\rm d}\chi}{{\rm d}\omega} = 
\eta \, \frac{kc\, \gamma}{|\Omega|^2}.
\end{eqnarray}
Since the medium is non-absorbing, high densities can be used and rather
small group velocities can be achieved. In fact values as low as 
few meters per second 
have been observed \cite{group}. 
In the region of linear dispersion 
the propagation equation for the slowly varying complex field amplitude
reads 
\be
\left(\frac{\partial}{\partial t}+v_{\rm gr}(z)
\frac{\partial}{\partial z}\right)
E(z,t)= 0
\ee
where we have taken into account a possible space dependence of the group
velocity. 
The solution of this equation
\be
E(z,t)= E
\biggl(0,t-\int^z_0{\rm d}z'\, \frac{1}{v_{\rm gr}(z')}\biggr),\label{sol-z}
\ee
describes slow propagation with an {\it invariant temporal pulse shape}.
Note, that the spatial shape is not conserved and depends 
on the spatial profile of $v_{\rm gr}(z)$.
The slow-down is a loss-less
linear process and hence all properties of the slowed pulse are conserved.
In particular also its quantum state.
The slow-down of the group velocity 
is thus interesting from the point of view of quantum memories as it
represents a temporary storage.

If a light pulse propagates in an 
absorption-free medium with linear dispersion, its total number of photons
is reduced by the ratio of the group velocity $v_{\rm gr}$ to the
vacuum speed of light
${n}/{n_0}={v_{\rm gr}}/{c}$.
However the time-integrated number of photons 
crossing a plane perpendicular to
the propagation direction is constant. Thus photons
must be temporarily stored in the combined system of atoms and control
field. 

The transfer mechanism between probe field and atomic system can 
be identified as STIRAP. 
To see this, let us consider a coherent probe pulse with all atoms
initially in the ground state $|b\rangle$.
This is a natural assumption since 
optical pumping will prepare the atoms in this state. 
At the initial time  the ground state $|b\rangle$
is then identical to the dark state
\begin{equation}
|d(z)\rangle = \cos\vartheta(z,t)\, |b\rangle -\sin\vartheta(z,t)\, |c\rangle
\enspace \longrightarrow\enspace |b\rangle
\end{equation}
where the mixing angle $\vartheta$ is now 
given by $\tan\vartheta(z,t) =\Omega_p(z,t)/\Omega$
with $\Omega_p(z,t)$ being the Rabi-frequency of the probe pulse. When the
front end of the 
probe pulse arrives at an atom, 
the mixing angle makes a small excursion from zero.
In this process a fraction of the state amplitude rotates
from $|b\rangle$ to $|c\rangle$ and 
energy is taken out of the probe pulse. When the probe pulse
reaches its maximum, the excursion of the mixing angle stops and
$\vartheta$ returns to zero. The state vector of the atoms
follows and  
rotates back to the ground state. Thus the entire energy is put back into 
the probe pulse at its back end. 
The slow-down of light in EIT can be understood as an
Raman adiabatic return of the atoms. 
In order for the process to be entirely adiabatic, the condition
\begin{equation}
\frac{\Omega^2}{\gamma} T_p \gg \sqrt{\eta k L }\label{adiabat_2}
\end{equation}
must be fulfilled, where $T_p$
is the pulse duration, $L$ the length of the medium, 
 and it was assumed that $|\Omega|\gg|\Omega_p|$. 
The presence of the factor $\sqrt{\eta k L}$ instead of
unity in (\ref{adiabat_2}), 
where $\alpha = \eta k L$ is the opacity
of the medium in the absence of EIT, is due to the fact that 
propagation effects need to be taken into account 
\cite{STIRAP,Fl-Manka96,EIT-spectr}.

\begin{figure}[ht]
\centerline{\epsfig{file=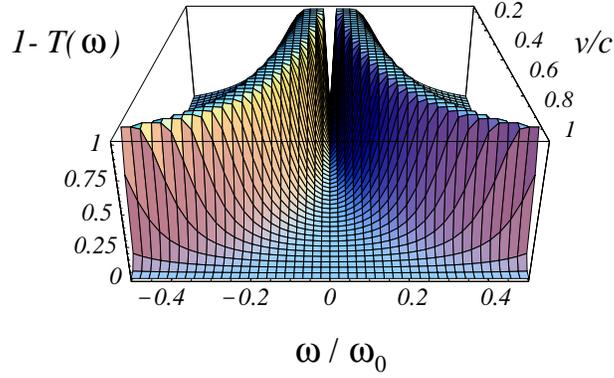,width=8.0cm}}
\vspace*{2ex}
\caption{
Transmission spectrum  
of EIT in units of $\omega_0= \eta k c$
as function of group velocity. When the transmission spectrum 
becomes narrower than the (constant) pulse spectrum,
strong absorption sets in. Parameters are
 $\alpha=\eta k L=20$, $\eta k c/\gamma=10$.
}
\label{EIT-slowdown}
\end{figure}

The inequality (\ref{adiabat_2}) has a very simple physical meaning. It 
states that the spectrum of the pulse should
be much less than the spectral window of transparency in EIT given by
\be
\Delta\omega_p\ll \Delta\omega_{tr}=\frac{|\Omega|^2}{\gamma}
\frac{1}{\sqrt{\eta kl}}. 
\label{trans_spectr}
\ee
It is instructive to express this condition in terms of the group
velocity. This yields
\be
\Delta\omega_p\ll \frac{v_{\rm gr}}{l}\, \sqrt{\eta k l}.
\ee
One immediately recognizes 
that a vanishing group velocity
implies a zero spectral width of transparency. 
This is illustrated in Fig.~\ref{EIT-slowdown} which shows the absorption
of a homogeneously broadened EIT medium as function of the 
detuning from resonance and the group velocity.

\

\

\begin{figure}[ht]
\centerline{\epsfig{file=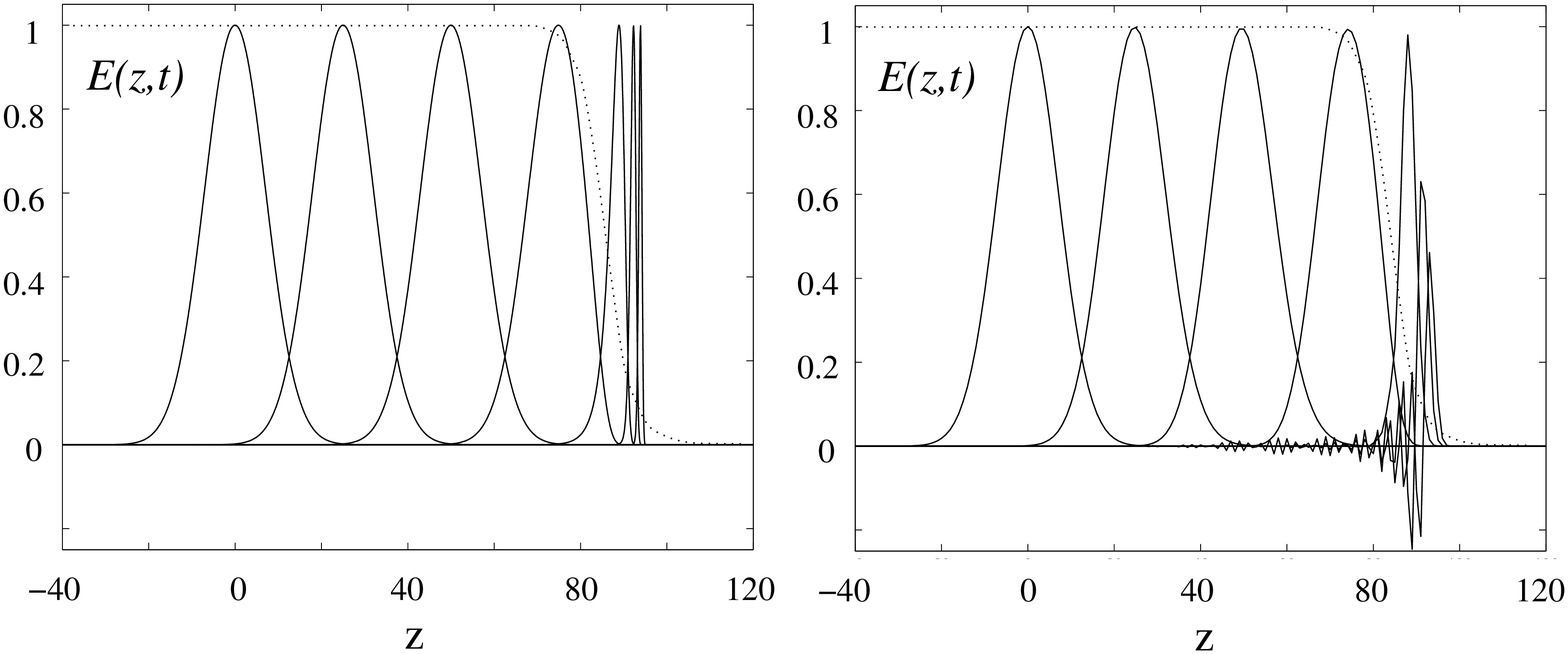,width=13.0cm}}
\vspace*{2ex}
\caption{
 probe pulse 
as it propagates into a soft ``road-block'', i.e. a region where 
$v_{\rm gr}(z)/c\to 0$, for times $t=0, 25, 50, 75, 100, 125, 150$;
{\it left:} if only linear dispersion
is taken into account, {\it right:} full numerical
solution of probe pulse propagation.  Axes are in arbitrary units with $c=1$.
The dotted line shows $v_{\rm gr}/c$. 
}
\label{1-d-EIT}
\end{figure}

Since EIT does not
change the spectrum of the probe pulse, it
is in this way not possible to bring a photon wave-packet to a complete stop. 
This is illustrated in Fig.~\ref{1-d-EIT}, where the propagation of a
pulse in an EIT medium with spatially decreasing group velocity is
depicted. The left curve shows the
effect of the linear dispersive part alone ($\chi^\prime$). Since the front
end of the pulse sees always a smaller group velocity than the back end, the
wave-packet becomes spatially compressed during the deceleration
according to $\Delta l(z)/\Delta l(0) = v_{\rm gr}(z)/v_{\rm gr}(0)$. 
The right curve shows for comparison a 
numerical solution of the 1-D propagation equations.
One clearly recognizes that at a particular position, where the pulse spectral
width becomes comparable to the decreasing transparency width, dissipation and
partial back reflection sets in.


\

Although EIT does not allow to bring a pulse to a complete stop, the
question arises whether it can be used as a {\it temporary}
memory.
Expressing the transparency width in terms of the
pulse delay time $\tau_d=n_g L/c$ for the medium yields
\be
\Delta\omega_{tr}=\sqrt{\eta k L }\, \frac{1}{\tau_d}.
\ee
Thus large delay times imply a narrow transparency window, which in turn 
requires a long pulse time. 
Hence there is an upper bound for the ratio of achievable delay (storage)
time to the initial pulse duration of a photon wave-packet
\be
\frac{\tau_d}{\tau_p} \le \sqrt{\eta k L }.
\ee
${\tau_d}/{\tau_p}$ is the figure of merit for any memory 
device. 
In practice, the achievable opacity $\alpha =\eta kL$ of atomic vapor systems 
is limited
to values below $10^4$ resulting in upper bounds for the 
ratio of time delay to pulse length of the order of 100.
Thus EIT media with ultra-small group velocity are only of 
limited use as temporary storage devices. 

The delay-time limitations are a common feature of any EIT-based
scheme which does not affect the spectrum of the probe pulse 
and sets strong limitations e.g. to freezing of light in moving media
\cite{Olga-slow} and so-called optical black holes based on EIT 
\cite{Leonhardt00}.
For instance a light pulse propagating in an EIT medium with a
vortex matter flow, as suggested in 
\cite{Leonhardt00}, will be absorbed or reflected before it reaches
the analog of the Schwarzschild-radius. 


\section{Light stopping by adiabatic rotation of dark-state polaritons}


\subsection{Time-dependent group velocity}


In the last section we have seen that the essential limitation of EIT
for a temporary memory and light stopping is the proportionality between
spectral transmission width and group velocity. 
Below a certain value of $v_{\rm gr}$
the transmission window becomes narrower than the pulse spectrum and the
pulse is absorbed unless its spectrum narrows as well. Changing the
spectrum of a pulse in a linear medium is only possible if the
medium properties change in time. Let us thus consider a hypothetical medium
with a time-dependent group velocity. The corresponding wave equation
reads
\be
\left(\frac{\partial}{\partial t}+v_{\rm gr}(t)
\frac{\partial}{\partial z}\right)
E(z,t)= 0.\label{E_propag_2}
\ee
It has the solution
\be
E(z,t)= E\biggl(z-
\int^t_0\!\!\!{\rm d}\tau
v_{\rm gr}(\tau),0\biggr),
\ee
which in contrast to the case of $v_{\rm gr}=v_{\rm gr}(z)$
describes a {\it spatially invariant} propagation.
At the same time the temporal profile is however not conserved and 
the spectrum of the probe field
changes during propagation. Assuming that $v_{\rm gr}$ changes only
slow compared to the field amplitude, one finds 
that the spectral width
narrows (broadens) according to 
\be
\Delta\omega_p(t)\approx \Delta\omega_p(0) \, 
\frac{v_{\rm gr}(t)}{v_{\rm gr}(0)}.
\ee
In this way
the bandwidth limitations of EIT can be overcome.
The spectrum of the probe pulse narrows in the same way as the
transparency width when the group velocity is reduced.
It is thus worthwhile
to consider the propagation of a probe pulse in an EIT medium
with an explicitly time dependent control field. 


\subsection{Polariton picture of EIT and ``stopping'' of light}


In order to describe the propagation of a weak probe pulse in an explicitly
time-dependent EIT
medium, let us consider  a one-dimensional 
model. Two fields, a strong and undepleted Stokes field
of Rabi-frequency $\Omega(t)$ and a weak probe pulse with
dimensionless field amplitude ${\cal E}(z,t)$ couple resonantly the
transitions in a 3-level medium as shown in Fig.~2. 
The propagation equation can be written in the
form
\be
\left(\frac{\partial}{\partial t}+
c\frac{\partial}{\partial z}\right){\cal E}
= { i} g N\, \rho_{ab}(z,t).\label{field}
\ee
where $g$ is the 
vacuum Rabi-frequency in an interaction volume $V$, with
$g^2 N = \eta k c \gamma$,  
$N$ being the number of atoms
in $V$. $\rho_{ab}(z;t)$ is the density matrix element of the atoms 
between the excited and ground states at position
$z$. 
Without the probe pulse, the atoms are assumed to be in state $|b\rangle$,
e.g. as a result of optical pumping by the Stokes field.
I.e. in zeroth order of the probe field one has $\rho_{bb}=1$ and all other
elements vanish. In first order of ${\cal E}$ the following 
relevant density-matrix equations are obtained:
\be
\dot\rho_{ab}
&=&-\gamma \rho_{ab} + i g {\cal E} + i\Omega
\rho_{cb},\\
\dot\rho_{cb}&=& i \Omega\, \rho_{ab}.
\ee
It is now convenient to introduce two new dimensionless field
variables corresponding to quasi-particles, called dark- and
bright-state polaritons \cite{FL-01}
\be
\Psi(z,t) &=&
\cos\theta(t)\, {\cal E}(z,t) - \sin\theta(t)\, \sqrt{N}\,
\rho_{cb}(z;t),\\
\Phi(z,t) &=&
\sin\theta(t)\, {\cal E}(z,t) + \cos\theta(t)\, \sqrt{N}\,
\rho_{cb}(z;t),
\ee
where $\tan^2\theta = g^2 N/\Omega^2=n_{\rm g}$. The group velocity
of slow-light propagation is thus given by $v_{\rm gr}=c\,\cos^2\theta$.
One can transform the equations of motion for the electric field and
the atomic variables into the new variables. This yields
\be
\biggl[\frac{\partial}{\partial t} +c\cos^2\theta
\frac{\partial}{\partial z}\biggr]\,
\Psi =-\dot\theta\Phi
-\sin\theta
\cos\theta\, c\frac{\partial}{\partial z}\Phi,\label{Psi-full}
\ee
and 
\be 
\Phi &=&\frac{\sin\theta}{g^2 N}
\biggl(\frac{\partial}{\partial t}+\gamma\Bigr)\Bigl(\tan\theta 
\frac{\partial}{\partial t}\biggr)
\Bigl(\sin\theta\,\Psi-\cos\theta\, \Phi\Bigr).\label{Phi-full}
\ee

Introducing an adiabaticity 
parameter $\varepsilon\equiv \bigl(g\sqrt{N} 
T\bigr)^{-1}$ with $T$ being a characteristic time, one can 
expand the equations of motion in a power series in $\varepsilon$.
In lowest order i.e. in the
adiabatic limit one finds $\Phi\approx 0$ and thus
the electric field amplitude as well as the spin coherence
are entirely determined by the amplitude of the dark-state
polariton
\be
{\cal E} = \enspace\cos\theta\, \Psi,\qquad {\rm and}\qquad
\sqrt{N} \rho_{cb} = -\sin\theta\, \Psi.
\label{sigma-Psi}
\ee
Furthermore all terms on the 
r.h.s. of eq.(\ref{Psi-full}) vanish and
one finds
\begin{eqnarray}
\left[\frac{\partial}{\partial t}+c\cos^2\theta(t)
\frac{\partial}{\partial z}\right]\Psi=0.\label{Psi-eq}
\end{eqnarray}
%
%


\subsection{``Stopping'' of light}


Eq.(\ref{Psi-eq}) describes a shape-preserving propagation with 
instantaneous velocity
$v=v_{\rm gr}(t)=c\cos^2\theta(t)$: 
\begin{equation}
\Psi(z,t)=\Psi\biggl(z- c\int^t_0\!\!\!{\rm d}\tau
\cos^2\theta(\tau),0\biggr).
\label{sol}
\end{equation}
For $\theta\to 0$, i.e. for a strong external drive field $\Omega^2\gg g^2 N$,
the polariton has purely photonic character $\Psi ={\cal E}$
and the propagation velocity is that of the vacuum speed of light.
In the opposite limit of a weak drive field $\Omega^2\ll g^2 N$ such that
$\theta\to \pi/2$, the polariton becomes spin-wave like
$\Psi =-\sqrt{N}\rho_{cb}$ and its
propagation velocity approaches zero. Thus the following mapping can be 
realized
\be
{\cal E}(z) \, \longleftrightarrow \, \rho_{cb}(z')\label{map}
\ee
with $z'=z+z_0=z+\int^\infty_0\!\!{\rm d}\tau c\cos^2\theta(\tau)$. 

Eq.(\ref{map})
is the essence of the transfer technique of quantum states from 
photon wave-packets propagating at the speed of light
to stationary atomic excitations 
(stationary spin waves). Adiabatically rotating the mixing angle from
$\theta=0$ to $\theta=\pi/2$ decelerates the polariton to a full stop,
changing its character from purely electromagnetic to purely atomic.
Due to the linearity of eq.(\ref{Psi-eq}), 
the quantum state of the polariton is not changed
during this process. 

\begin{figure}[ht]
\centerline{\epsfig{file=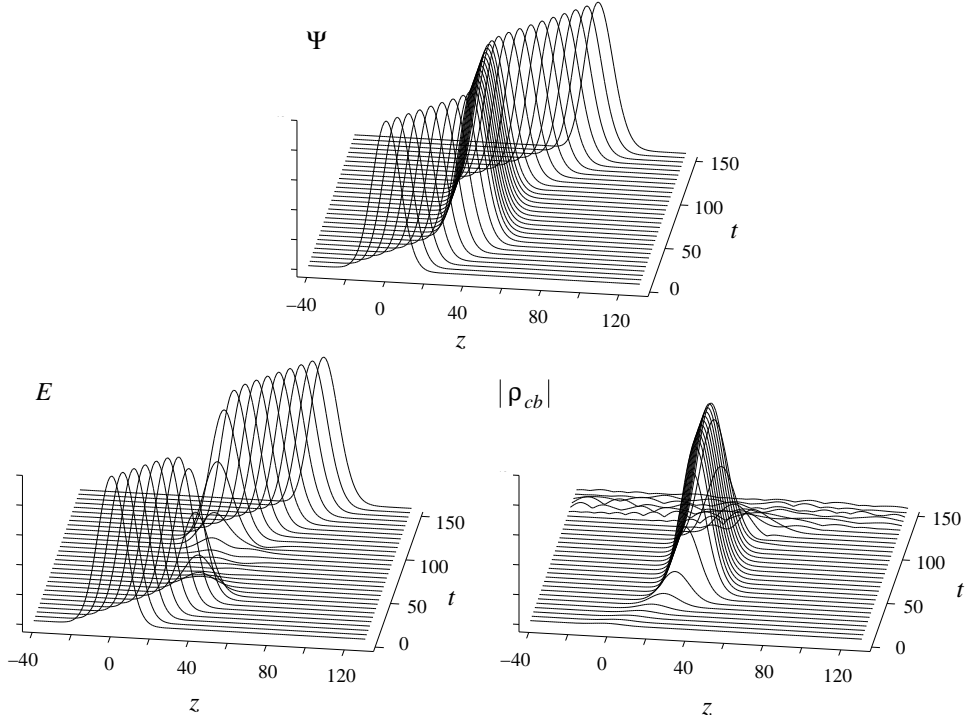,width=13.0cm}}
 \vspace*{2ex}
 \caption{Numerical simulation of dark-state polariton propagation 
with envelope $\exp\{-(z/10)^2\}$. 
The mixing angle is rotated from $0$ to $\pi/2$ and back
according to $\cot\theta(t)  =$ $ 100(1-0.5 \tanh[0.1(t-15)]$ $
 + 0.5\tanh[0.1(t-125)])$. 
{\it top:} polariton amplitude, {\it bottom-left:} electric field amplitude
, {\it bottom-right:} atomic spin coherence; all in arbitrary units.
 Axes are in arbitrary units with $c=1$. }
\label{stop}
\end{figure}

Likewise the polariton can be
re-accelerated to the vacuum speed of light; in this process the stored 
quantum state is transferred back to the field. 
This is illustrated in Fig.\ref{stop}, where we have shown 
the coherent amplitude of a dark-state polariton which results from an initial
light pulse as well as the corresponding  electromagnetic and matter
components obtained from a numerical solution of the 1-D Maxwell-Bloch
equations. 

\

\begin{figure}[ht]
\centerline{\epsfig{file=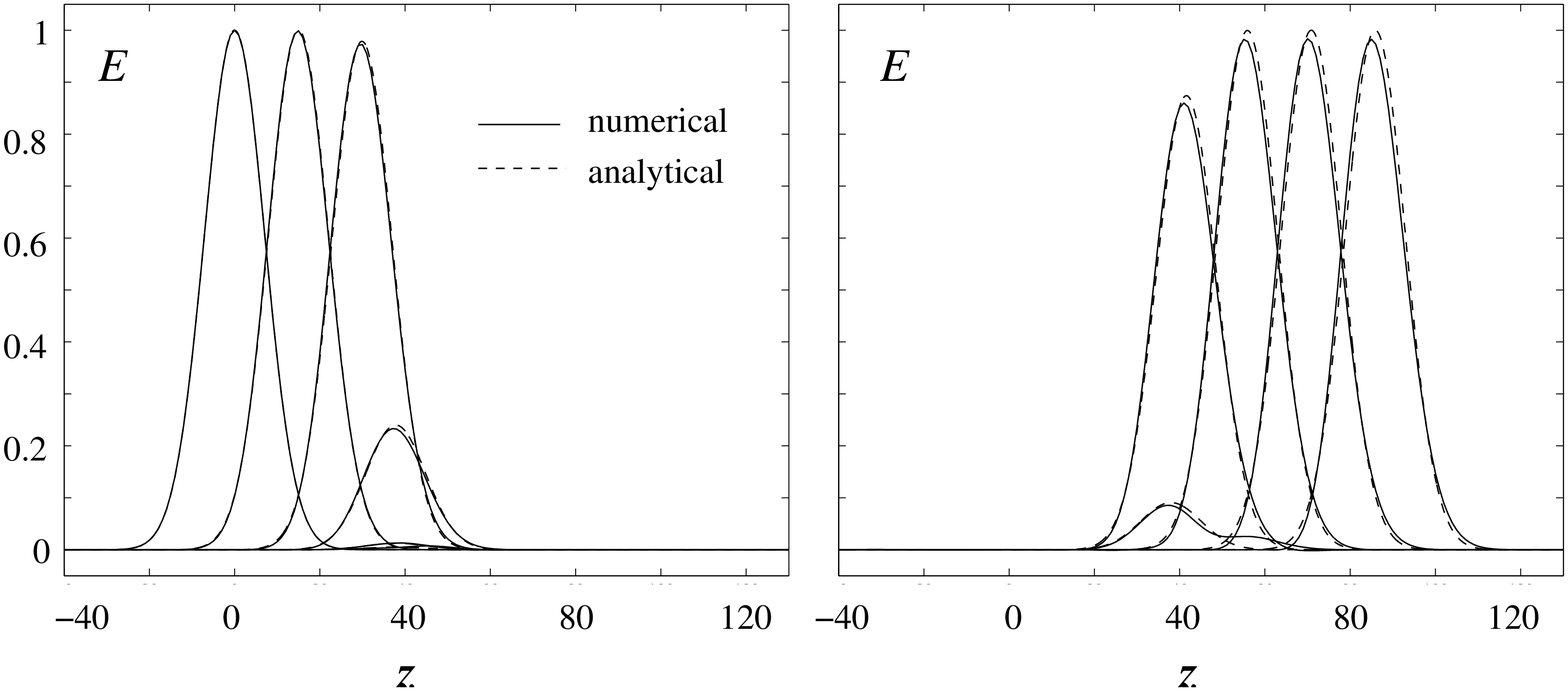,width=12.0cm}}
 \vspace*{2ex}
 \caption{Comparison of numerical (full line) and 
approximate analytical
results (dashed line) following from eq.(\ref{sol}) for example of
Fig.~\ref{stop}. {\it left:} electric field amplitude in arbitrary units
for $t=0, 15, 30, 45, 60$. {\it right:} the same for
$t=90,105,120,135,150$}
\label{fig-compare}
\end{figure}

As noted above, the spectrum of the polariton and thus of the 
probe pulse narrows in the same way as the EIT transmission window
when the group velocity is reduced as a function of time.
The adiabatic slow-down of a dark-state polariton thus avoids the
bandwidth limitation of EIT and a light pulse can indeed be brought to
a full stop. This is the essential difference of the present scheme to
the light-freezing proposal of Ref.~\cite{Olga-slow}. 
In the adiabatic transfer process all information carried by the pulse, i.e.
also its quantum state, are transferred to a collective 
Raman or spin excitation of atoms. 

In order to prove the validity of the simple analytic solution (\ref{sol}),
we have compared
in Fig.~\ref{fig-compare} exact numerical and analytical results for the
situation of Fig.~\ref{stop}. As can be recognized there is an excellent 
agreement.


\subsection{Adiabatic condition and non-adiabatic corrections}


The analysis of the conditions for Raman adiabatic passage in single-atom
cavity QED in sec.~2 lead to the experimentally challenging strong-coupling 
requirement. Let us now investigate the corresponding
requirements for the present scheme. 
To this end we take into account the first-order correction
to the adiabatic solution (\ref{sol}). 
Making use of the zeroth-order relation
$\dot\Psi=-c \cos^2\theta\partial\Psi/\partial z$ to eliminate time 
derivatives in favor of spatial derivatives we find
\be
&&\left[\frac{\partial}{\partial t}+c\cos^2\theta
\frac{\partial}{\partial z}\right]\Psi =  -A(t)\Psi
+B(t)c\frac{\partial}{\partial z}\Psi
+\, C(t)c^2  \frac{\partial^2}{\partial z^2}\Psi
- D(t)c^3 \frac{\partial^3}{\partial z^3}\Psi,\label{Psi_correct}
\ee
where
\be
A(t) &=& \Bigl(\gamma+\frac{1}{2}\frac{\partial}{\partial t}\Bigr)
\left(\frac{\dot\theta^2\sin^2\theta}{g^2 N}\right),\qquad
B(t)= \frac{\sin\theta}{3 g^2 N} \frac{\partial^2}{\partial t^2} \sin^3\theta,\\
C(t) &=&  \Bigl(\gamma+\frac{1}{2}\frac{\partial}{\partial t}\Bigr)
\frac{\sin^4\theta\cos^2\theta}{g^2 N},\qquad\enspace
D(t) = \frac{\sin^4\theta\cos^4\theta}{g^2 N}.
\ee
$A(t)$ describes homogeneous losses due to 
non-adiabatic transitions followed by spontaneous emission.
$B(t)$ gives rise to a correction
of the polariton propagation velocity, $C(t)$ results in a pulse spreading 
by dissipation of high spatial frequency components and $D(t)$ leads to
a deformation of the polariton. 
Since all coefficients depend only on time, eq.(\ref{Psi_correct}) can
be solved by spatial Fourier transform and simple integration in time.
In order to neglect dissipation, the terms containing $A(t)$ and $C(t)$
must be small, which results in the two conditions
\be
\frac{\gamma c^2}{L_p^2} \int_0^\infty\!\!{\rm d}t\,
\frac{\cos^2\theta(t)}{g^2N}=\frac{\gamma c L}{g^2 N L_p^2}\ll 1,
\label{ad-cond1}
\ee
and 
\be
\gamma\int_0^\infty\!\!\!{\rm d}t\, \frac{\dot\theta^2\sin^2\theta}{g^2 N}
=\gamma\int_0^\infty\!\!\!{\rm d}t\, \frac{\dot\theta^2}{g^2 N+\Omega^2(t)}
\ll 1.
\label{adb}
\ee
To simplify the first expression we 
have used here the approximation $\sin\theta\approx 1$.
$L_p$ denotes the pulse length in the medium at the initial time.

The first condition should be compared and contrasted to the strong
coupling condition of sec.~2. Using $T_p=L_p/c$ one finds
\be
\frac{g^2 N^\prime}{\gamma} T_p \gg 1  \qquad\longleftrightarrow
\qquad \frac{g^2 }{\gamma} T_p \gg 1, 
\ee
where $N^\prime \equiv N L/L_p$ is the number of atoms in a volume
of length $L_p$ (rather than $L$). The large factor $N^\prime$
is a signature of the collective interaction and strongly
alleviates the requirements in the many-atom system as compared to
the single-atom one.
Condition (\ref{ad-cond1}) has again a simple physical meaning.
It states that the {\it initial} pulse spectral width
has to be much less than the {\it initial} transparency width, i.e.
at the time when the pulse enters the medium. 
\be
\Delta \omega_{p}(0) \ll \Delta \omega_{tr}(0). 
\ee
This is easily satisfied, if the probe pulse is not too
short. Practically it does prevent however light storage of pulses 
shorter than a few nanoseconds. 

\begin{figure}[ht]
\centerline{\epsfig{file=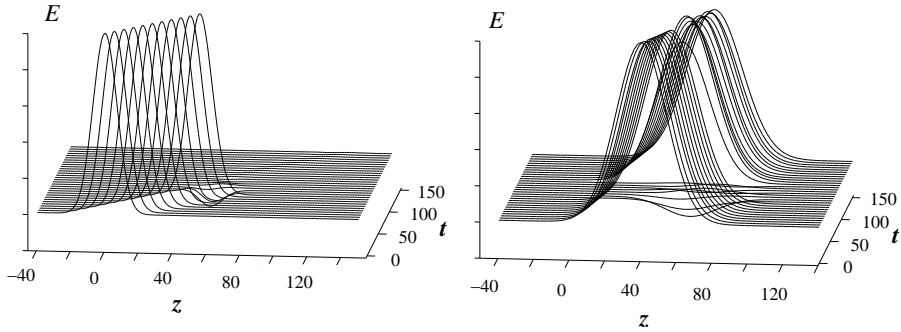,width=12.0cm}}
 \vspace*{2ex}
 \caption{propagation of light pulse in EIT medium with sudden
switching of $\cos\theta$. {\it left:} $\cos\theta$ is switched from
1 to zero at $t=50$ and back at $t=90$, {\it right:} same switch, but with
initial and final values of $\cos^2\theta(\pm \infty)=0.1$, only 
electromagnetic component of the polariton (here 10\%) gets lost;
$\gamma=1$.
}
\label{switch-1}
\end{figure}

The second condition is
well-known from adiabatic  passage 
\cite{Fl-Manka96,Vitanov} and sets a limit to the 
rotation velocity $\dot\theta$ of the mixing angle and hence to
the deceleration/acceleration of the polariton.
In order to stay in the adiabatic regime at all times the
characteristic time scale $T$ has to fulfill
the condition
\be
T \gg {l_{abs} \over c} {v_{\rm gr}^0 \over c},
\ee
where $l_{abs} = c \gamma/g^2 N$ is the absorption length 
in the absence of EIT.
We note that for realistic experimental parameters the quantity 
on the right hand  side is extremely small on all relevant time scales. 
It should be noted that a sudden and sharp switch-off of the 
drive field leads to a complete loss of the entire electromagnetic 
component of the polariton.
This is illustrated by numerical simulations in Fig.\ref{switch-1}. 
The left figure shows 
the propagation of a light pulse in an EIT medium, where the
cosine of the mixing angle is switched from 1 to 0 at $t=50$ and
back to 1 at $t=90$. Apart from some small and rapidly decaying
Rabi-oscillations, the light field is immediately absorbed and cannot
be regenerated. 
On the other hand
the losses are small 
if the cosine of the mixing angle is initially 
significantly less than unity. 
This is shown in the right picture, where $v_{\rm gr}/c=
\cos^2\theta(\pm\infty)=0.1$
and hence the output amplitude is only reduced by 10\%.

\

\section{Quantum memory and decoherence}


An important question when discussing the application of the 
many-atom system to a quantum memory is the sensitivity to errors.
Thus in the following the influence of decoherence processes on the
fidelity of the state storage will be discussed.

The most essential properties of a 
quantum memory based on collective excitations can be
understood looking at the case of a single mode of the radiation field
as realized e.g. in a single mode optical cavity. 
Consider a collection of $N$ 3-level atoms 
as shown in Fig.~\ref{3-level}.
The dynamics of 
this system is described by the complex Hamiltonian ($E_b=\hbar\omega_b=0$):
\be
H &=&\hbar\omega a^\dagger a
+\hbar(\omega_a -i\gamma)\sum_{j=1}^N \sigma_{aa}^j +\hbar\omega_c 
\sum_{j=1}^N \sigma_{cc}^j 
+\label{ham}\\
&&+
 \hbar g \sum_{i = 1}^N   a\sigma_{ab}^i + 
\hbar\Omega(t) {\rm e}^{-i\nu t}
\sum_{i = 1}^N  
\sigma_{ac}^i + {\rm h.c.} .  \nn
\ee
Here $\sigma_{\mu\nu}^i = |\mu\rangle_{ii}\langle \nu|$ is the 
flip operator of the $i$th atom and the vacuum Rabi-frequency
is assumed to be equal for all atoms. We also have introduced an imaginary
part to the Hamiltonian to take into account losses from the 
excited state.

\begin{figure}[ht]
\centerline{\epsfig{file=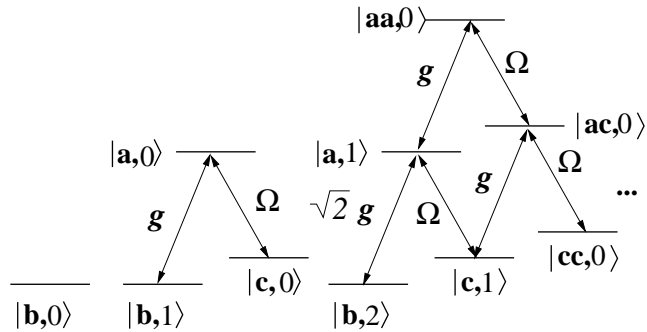,width=8.5 cm}}
\vspace*{2ex}
\caption{
Coupling of
bare eigenstates of atom + cavity system for at most two photons
}
\label{3-level-collective}
\end{figure}

When all atoms are prepared initially in  level $|b\rangle$ 
the only states coupled by the interaction are the totally symmetric
Dicke-states \cite{Dicke54}
\be
|{\bf b}\rangle_N &=& |b_1,b_2,\dots,b_N\rangle,\\
|{\bf a}\rangle_N &=& \frac{1}{\sqrt{N}} \sum_{j=1}^N
|b_1,\dots,a_j,\dots,b_N\rangle,\\
|{\bf c}\rangle_N &=& \frac{1}{\sqrt{N}} \sum_{j=1}^N
|b_1,\dots,c_j,\dots,b_N\rangle,\\
|{\bf aa}\rangle_N &=& \frac{1}{\sqrt{2N(N-1)}}\sum_{i\ne j=1}^N
|b_1,\dots,a_i,\dots,a_j,\dots,b_N\rangle,\quad{\rm etc.} .
\ee
The couplings within the sub-systems corresponding to 
a single and a double excitation 
are shown in  Fig.~\ref{3-level-collective}.
 

The interaction of the $N$-atom system with the quantized
radiation mode has now a family of dark-states
\be
|D,n\rangle_N &=& 
\sum_{k=0}^n\sqrt{\frac{n!}{k!(n-k)!}}
(-\sin\theta)^k (\cos\theta)^{n-k}
|{\bf c}^k,n-k\rangle_N,\label{darkn}\\
&& \qquad\tan\theta(t) \equiv\frac{g\sqrt{N}}{\Omega(t)}.\nn
\ee
It should  
be noted that although the dark states $|D,n\rangle_N$ are degenerate
they belong to exactly decoupled sub-systems. 
This means there is no transition between them even if 
non-adiabatic corrections are taken into account.
Adiabatically rotating the mixing angle $\theta$ from $0$ to 
$\pi/2$ leads to a complete and reversible transfer of all 
photonic states to a collective atomic
excitation if the maximum number of photons $n$ is less than the number
of atoms  $N$. 
If the initial quantum state of the single-mode light field 
is in a mixed state described by the density matrix 
$\hat\rho_f=\sum_{n,m} \rho_{nm}\, |n\rangle\langle m|$, the 
transfer process generates a quantum state of collective excitations
according to
\be
\sum_{n,m}\rho_{nm} \, |n\rangle\langle m|
 \otimes |{\bf b}\rangle_{N\, N}\langle{\bf b}|
\quad\longleftrightarrow\quad |0\rangle\langle 0|\otimes 
\sum_{n,m} \rho_{nm}\, |{\bf c}^n\rangle_{N\, N}\langle {\bf c}^m|.
\ee

Thus the storage state corresponding to a photon Fock state $|n\rangle$
is an entangled many-particle state $|{\bf c}^n\rangle_N$. These states
are known to be 
highly susceptible to decoherence. For example if any one of the
$N$ atoms undergoes a spin flip, corresponding to $|b\rangle_j
\leftrightarrow |c\rangle_j$, an orthogonal state is created. When the
probability of a single-atom error is $p$, the total
probability that an orthogonal state is created is $P=1-(1-p)^N
\sim Np$. Thus one might conclude that collective quantum memories
display a substantially enhanced rate of decoherence, which is a substantial
drawback that out-weights the advantage of 
ensembles for the coupling of light and matter.
Although the following discussion is
far from complete, some arguments why this
is indeed not the case will be presented now \cite{decoh}. 

For this the system Hamiltonian will first be brought into a more 
appropriate form. Separating the oscillatory factor ${\rm e}^{-i\nu t}$
by a canonical transformation, assuming two
photon resonance,
i.e. $\omega=\omega_c+\nu$, 
 and adiabatically eliminating the
excited state $|a\rangle$ yields 
\be
H=\hbar\omega \Bigl(
\Psi^\dagger \Psi+\sum_l\Phi_l^\dagger\Phi_l\Bigr)
-i\hbar \frac{\Omega^2(t)}{\gamma} \sum_l \Phi_l^\dagger\Phi_l.
\label{ham-pol}
\ee
Here we have assumed adiabatic conditions
and introduced the dark-state polariton operator
\be
\Psi = \cos\theta(t)\, a -\sin\theta(t)\, \frac{1}{\sqrt{N}}
\sum_{j=1}^N \sigma_{bc}^j,\label{Psi-def}
\ee
as well as the $N$ bright-state polariton operators
\be
 \Phi_0 &=& \sin\theta(t)\, a +\cos\theta(t)\, \frac{1}{\sqrt{N}}
\sum_{j=1}^N \sigma_{bc}^j,\label{Phi0-def}\\
\Phi_l &=&  
\frac{1}{\sqrt{N}}
\sum_{j=1}^{N} \sigma_{bc}^j \, \exp\left\{2 \pi i \frac{l j}{N}\right\}
\qquad {\rm for}\quad l\in\{1,2,\dots,N-1\}.\label{Phi-def}
\ee
Expression (\ref{ham-pol}) shows that under adiabatic conditions
all excitations of the dark-state polariton $\bigl(\Psi^\dagger\bigr)^n\,
|{\bf b},0\rangle_N $
are conserved, 
while all bright-polariton excitations decay via
optical pumping into the excited state and subsequent 
spontaneous emission. 

It can also be seen from eqs.(\ref{ham-pol} - \ref{Phi-def})
that in the read-out process of the memory, where
$\theta:\, \pi/2\, \to\, 0$ only dark-polariton excitations
will be transferred to field excitations. Thus from the
point of view of photon storage all
collective states with the same number of dark-polariton
excitations are equivalent. For example the symmetric 
dark-state
\be
|D,n\rangle_N = \frac{1}{\sqrt{n!}} \Bigl(\Psi^\dagger\Bigr)^n
|{\bf b},0\rangle_N,
\ee
which is a storage state of a radiation-mode Fock-state with $n$ photons
is equivalent to all states with arbitrary number of bright-polariton
excitations as long as there are 
exact $n$ excitations of $\Psi$:
\be
|D,n\rangle_N \quad \hat = \quad
\bigl(\Phi_1^{(\dagger)}\bigr)^k 
\dots \bigl(\Phi_2^{(\dagger)}\bigr)^l 
\bigl(\Psi^\dagger\bigr)^n
|{\bf b},0\rangle_N.
\ee
The existence of these {\it equivalence classes} of storage states
has an important consequence.
All transitions within an equivalence class of states 
caused by environmental interactions 
leave the fidelity of the quantum memory unaffected.
This
leads to an exact compensation of the enhanced probability
of a single-atom error to occur.

To illustrate this, let us consider the case of a Fock state
of the radiation field $|n\rangle$ stored in the
symmetric dark-state $|D,n\rangle_N$. A random spin flip
of atom $j$ from state $|b\rangle$ to $|c\rangle$ 
can be described by the operation
$|D,1\rangle_N \rightarrow \sigma_{cb}^j
|D,1\rangle_N$.
Assuming $\theta=\pi/2$ and making use of
\be
\sigma_{cb}^j
=\frac{1}{\sqrt{N}} \left(
\sum_{l=1}^{N-1} \exp\left\{-2\pi i\frac{lj}{N}\right\}
\, \Phi_l^\dagger +\Psi^\dagger\right)
\ee
one finds
\be
|D,n\rangle\enspace \longrightarrow \enspace 
\sum_{j=l}^{N-1}\frac{\exp\left\{-2\pi i\frac{lj}{N}\right\}}
{\sqrt{N}}\, \Phi_l^\dagger |D,n\rangle
+ \frac{1}{\sqrt{N}} \Psi^\dagger|D,n\rangle.
\ee
Only the last component belongs to a state that is not in the same
equivalence class as $|D,n\rangle_N$. Thus the 
probability to leave the
equivalence class is only $1/N$. This factor exactly compensates the
factor $N$ of the total probability of a spin flip error to occur in
any one atom. 

One can show that similar arguments hold for all possible
non-cooperative decoherence mechanisms. Hence the 
sensitivity of the collective quantum memory does not depend on the number
of particles involved. The sensitivity of 
many-particle quantum memories to decoherence is not enhanced
as compared to the single-atom counterparts, given that there are
no coopertive decay processes. 


\section{Summary}


A technique for a controlled and coherent transfer of
the quantum state of photons to and from  collective spin excitations
of ensemble of atoms has been discussed. The underlying mechanism is
an adiabatic rotation of a dark-state polariton between a purely
photonic and a purely atomic excitation. Due to the collectively
enhanced interaction cross section, the strong-coupling requirement
of single-atom cavity QED is eliminated.
The mapping technique,
although superficially similar looking, is quite different from
conventional methods of optical data storage based on
optical pumping. In the present set-up no spontaneous emission 
 is present,
making a reversible storage on the level on individual light quanta possible. 

It was shown that despite the entangled character of the storage states,
the collective quantum memory is not more sensitive to decoherence than
a single-particle system. This is due to the existence of
equivalence classes of many-particle excitations representing the
same stored quantum state of light. 

While many-particle excitations may have potential
advantages over single-atom ones with respect to 
storage purposes, processing of the information
in collective and thus delocalized qubits is more difficult.
Recently a method was suggested, however, to implement 
quantum logic operations between collective excitations \cite{dipole}.
The underlying physical mechanism is here a blockade effect caused
by resonant dipole-dipole interactions in Rydberg states which is
not sensitive to the actual separation between two atoms.


\section*{Acknowledgement}


The authors would like to thank M.D. Lukin, L.M. Duang, J.I. Cirac and P. Zoller
for stimulating discussions. The financial support of the Deutsche Forschungsgemeinschaft
under contract no. FL210/10 is gratefully acknowledged.


\def\etal{\textit{et al.}}

\end{document}